\documentclass[conference]
{IEEEtran}

\usepackage[square,numbers]{natbib}

\usepackage{amsmath,amssymb,amsfonts}
\usepackage{algorithmic}
\usepackage{graphicx}
\usepackage{float}
\usepackage{color, soul}

\graphicspath{ {./Images/} }
\usepackage{textcomp}
\usepackage{xcolor}
\usepackage{caption}

\usepackage[colorlinks=False]
{hyperref}
\IEEEoverridecommandlockouts
\def\BibTeX{{
\rm B\kern-.05em{\sc i\kern-.025em b}\kern-.08em
    T\kern-.1667em\lower.7ex\hbox{E}\kern-.125emX}}

\begin{document}

\title{Arabic Music Classification and Generation using Deep Learning}

\author{\IEEEauthorblockN{Mohamed Elshaarawy$^*$, Ashrakat Saeed$^*$, Mariam Sheta$^*$, Abdelrahman Said$^*$, Asem Bakr$^*$, Omar Bahaa$
^*$, Walid Gomaa$^
{*,\dagger}$}
\IEEEauthorblockA{$^*$Egypt-Japan University of Science and Technology, Alexandria, Egypt.}
\IEEEauthorblockA{$^{\dagger}$Faculty of Engineering, Alexandria University, Alexandria, Egypt.}
\IEEEauthorblockA{\{mohamed.elshaarawy, ashrakat.saeed, maryem.abousaad,abdelrahman.said, asem.abdelhamid, omar.bahaa, walid.gomaa\}@ejust.edu.eg}}

\maketitle

\begin{abstract}
\par This paper proposes a machine learning approach for classifying classical and new Egyptian music by composer and generating new similar music. The proposed system utilizes a convolutional neural network (CNN) for classification and a CNN autoencoder for generation. The dataset used in this project consists of new and classical Egyptian music pieces composed by different composers.

\par To classify the music by composer, each sample is normalized and transformed into a mel spectrogram. The CNN model is trained on the dataset using the mel spectrograms as input features and the composer labels as output classes. The model achieves 81.4\% accuracy  in classifying the music by composer, demonstrating the effectiveness of the proposed approach.

\par To generate new music similar to the original pieces, a CNN autoencoder is trained on a similar dataset. The model is trained to encode the mel spectrograms of the original pieces into a lower-dimensional latent space and then decode them back into the original mel spectrogram. The generated music is produced by sampling from the latent space and decoding the samples back into mel spectrograms, which are then transformed into audio.

 In conclusion, the proposed system provides a promising
 approach to classifying and generating classical Egyptian music,
 which can be applied in various musical applications, such as
 music recommendation systems, music production, and music
 education.

\end{abstract}


\maketitle

\section{Introduction}
\label{sec: Introduction}
This project aims to leverage deep learning techniques for the classification and generation of Arabic music. The project encompasses several phases, starting with an extensive data collection process. A dataset comprising music files from various genres was assembled. Subsequently, rigorous pre-processing techniques were employed to optimize the data for the subsequent modeling stages.
The pre-processing phase involved exploring different approaches to identify the most effective methods for enhancing model performance. This involved techniques such as audio normalization, feature extraction, and data augmentation. By carefully refining the pre-processing pipeline, the project sought to maximize the quality and relevance of the input data for subsequent modeling steps.
The next phase involved training and fitting deep learning models for both music classification and music generation tasks. The classification models were subjected to iterative testing and refinement to achieve satisfactory accuracy levels. For the music generation aspect, the fitted model was used to generate new musical compositions. Post-processing techniques were employed to refine and enhance the generated music, ensuring that it adhered to the desired musical characteristics and structure. These post-processing steps may have included melody harmonization, rhythm adjustment, and tempo normalization.
By employing advanced deep learning techniques and thorough pre-processing methodologies, this project aims to advance the field of Arabic music classification and generation. The results obtained from this research can potentially contribute to the broader domain of music analysis, synthesis, and creative expression.

\section{Related Work} 
\label{SecWork}
\par In recent years, there has been a growing interest in the field of music classification and generation. Several studies have explored various techniques and methodologies to address the challenges associated with this domain. In this section, the relevant literature on music composer classification, and generation highlighting the key approaches and achievements are reviewed.

\par One notable paper in this field is titled "Composer Classification Models for Music-Theory Building" by Herremans, Martens, and Sorensen (2015)\cite{comp}. In their study, the researchers adopted a data-driven approach by extensively analyzing a large database of existing music. The main objectives of their work were twofold: first, to develop an automated system capable of accurately distinguishing between compositions from different composers, and second, to identify the significant musical attributes that contribute to this distinction.

\par A recent Kaggle project by Holst\cite{kaggleProject} focused on music composer classification using a machine learning approach. The project involved building a model to classify musical compositions based on the underlying composer's style. Holst extracted various features from the audio recordings and employed a classification algorithm to differentiate between composers. The results obtained in the Kaggle project provide valuable insights into the application of machine learning techniques for music composer classification.

\par The paper titled "A Comprehensive Survey on Deep Music Generation"\cite{comprehensive} published in the Journal of Artificial Intelligence and Data Analytics is a valuable resource that we utilized to deepen our understanding of Variational Autoencoders (VAEs) and explore their potential for sound generation. We gained insights into the capabilities and limitations of VAEs in generating realistic sound samples. This knowledge served as a foundation for navigating further improvements in our own model, allowing us to incorporate the latest advancements and adapt them to the specific requirements of our Arabic music generation task.

\par The paper titled "An Introduction to Variational Autoencoders" by Diederik P. Kingma and Max Welling\cite{variational}, serves as a fundamental resource for understanding Variational Autoencoders (VAEs). It provides a comprehensive introduction to the theory and practical implementation of VAEs, exploring their underlying principles and applications in machine learning. This paper played a crucial role in our research as we leveraged its insights to comprehend the workings of VAEs and applied them to our own model for generating Arabic music. 
\section{Dataset}
\label{SecDataset}
\par Our classification and generation models were trained on a diverse data set of Arabic music. The data set was carefully curated to include a variety of genres, including classical, modern, piano, and oud songs, to ensure the model was exposed to a wide range of musical styles. The classical category consisted of compositions by renowned Egyptian artists such as Mohamed Abdelwahab, Reyad ElSonbaty, and Balegh Hamdy. The modern category included songs by popular composers like Waleed Saad, Mohamed Rahem, and others. Must mention that our data size was 256 Arabic songs(MUSIC ONLY), ranging from 2.5 minutes to 50 minutes in length.

\par Overall, the data set used for training the music generation model was carefully selected to provide a diverse set of musical styles and genres while maintaining consistency in length. The data set and preprocessing methods used in this study can serve as a foundation for future music generation research in Arabic music and related fields.

\section{Methodology} 
\label{SecMethod}

\subsection{Composers Classification Model}
\label{sec: classification}
\subsubsection{Preprocessing}
\label{sec: Preprocessing}

\par This study aimed to develop an optimal model for music classification, specifically in the context of composer identification. To achieve this goal, a thorough investigation was conducted through a series of five experiments. The ultimate objective was to identify and establish the most effective approach that yields accurate results in composer classification.

\par In the first experiment, A dataset comprising compositions from 11 different composers, with unevenly distributed data between them, representing diverse musical styles and genres, was selected.

\par To ensure consistency of results, a standardized preprocessing pipeline was followed. The steps in this pipeline included the conversion of the audio files into WAV format and segmenting and normalizing the compositions. Segmentation involved dividing the compositions into smaller segments of 10 seconds each, with a 2-second hop-length. Normalization was performed to ensure consistent volume levels across all segments.
\par The next step involved extracting the MFCC features from the normalized audio segments. This feature extraction technique captures essential acoustic characteristics of the audio signals. To achieve uniform dimensions across all segments, padding techniques were applied. The segments were padded to match a target shape of (20, 938), ensuring that the input data has the same dimensions, which is necessary for training the model consistently.

\begin{figure}[H]
    \centering
    \includegraphics[width=0.4\textwidth, height=4.5cm]{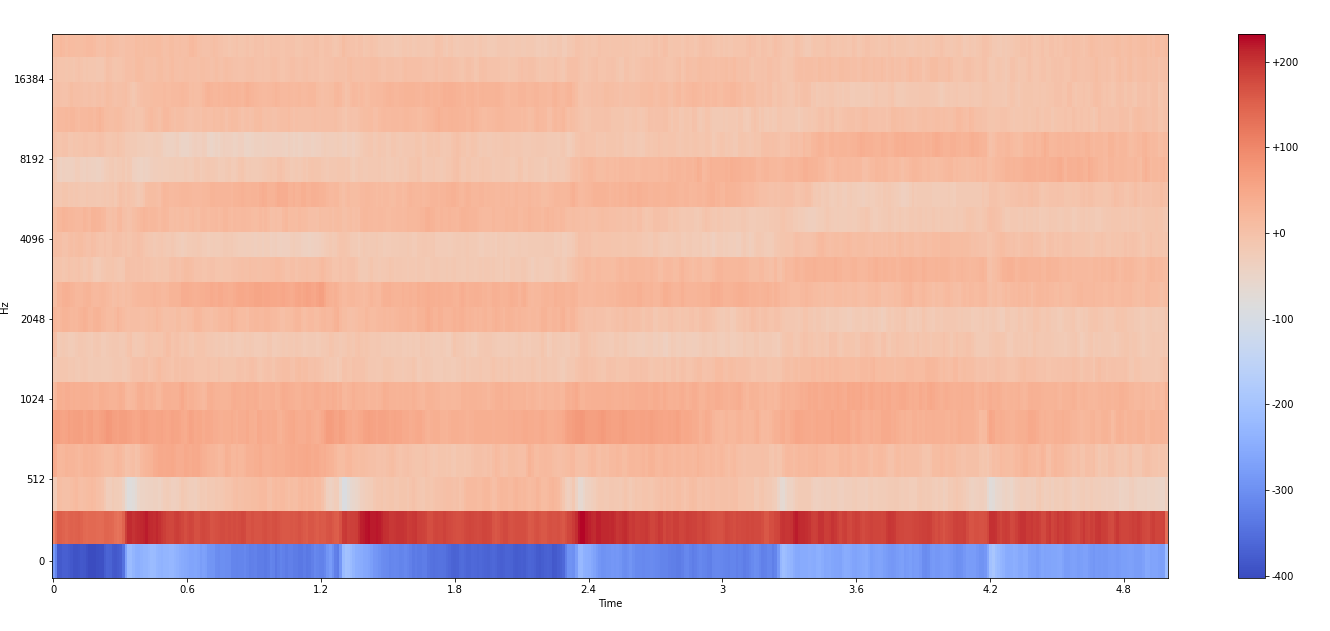}
    \caption{Example of generated MFCC}
    \label{fig:VAE2}
\end{figure}

\par The data was split into training, testing, and validation sets using a ratio of 70:15:15, respectively. The training set, comprising 70\% of the data, was used to train the model and optimize its parameters. The testing set, representing 15\% of the data, served as an independent dataset to assess the model's ability to generalize to unseen examples. The validation set, also consisting of 15\% of the data, played a vital role in monitoring the model's performance during training and preventing overfitting.

\par The training set (X-train) consisted of the preprocessed audio segments, while the corresponding target labels (Y-train) indicated the composer for each segment. Separate testing sets (x-test and y-test) were created to evaluate the model's performance on unseen data. Additionally, a validation set (x-val and x-val) was formed by selecting a portion of the unseen data through training to monitor the model's performance during training.

\par In the second experiment, a different approach was taken to evenly distribute the data among the composers. Instead of including all 11 composers, a subset of 9 composers was selected based on the availability of a sufficient number of data samples for each composer. This ensured a more balanced dataset for analysis.

\par
\begin{figure}[H]
    \centering
    \includegraphics[scale=0.5]{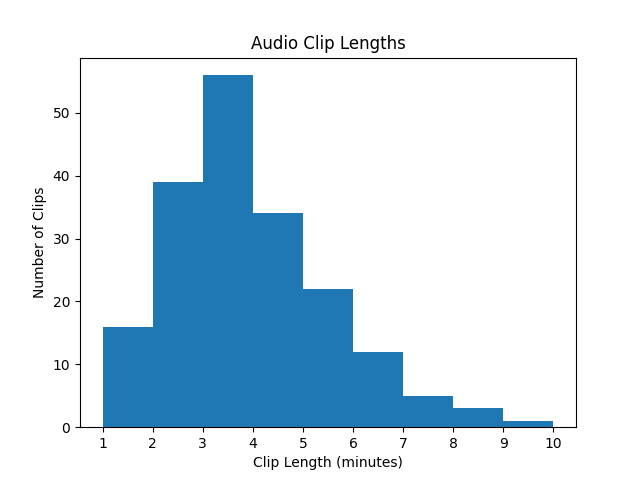}
    \caption{Audio Clip Lengths}
    \label{fig:clipLength}
\end{figure}

 To gain further insights into the dataset, the average duration of each audio clip was calculated, as shown in figure \ref{fig:clipLength}. Additionally, it was decided how many clips should be taken from each composer.

\par
\begin{figure}[H]
    \centering
    \includegraphics[width=.4\textwidth, height=5cm]{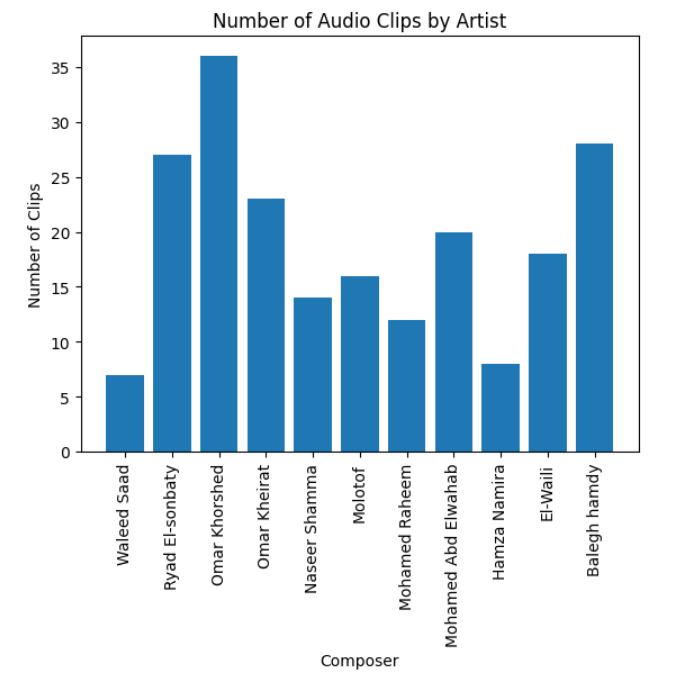}
    \caption{Number of Clips}
    \label{fig:clipsNumber}
\end{figure}

 Similarly, the number of clips available for each composer in the dataset was examined, as shown in figure \ref{fig:clipsNumber}. By considering the distribution of clips per composer, the dataset's balance could be evaluated and any significant variations in the amount of data samples across composers could be identified.

\par After calculating the number of clips and segments, several decisions were made regarding the dataset for the second experiment. 

\begin{enumerate}
    \item Maximum of 20 clips for each Composer: To maintain consistency and fairness among the composers. This ensured that no composer had an overwhelming number of clips compared to others.
    \item Elimination of Hamza Namira and Waleed Saad data: In the interest of maintaining a balanced dataset, it was decided to exclude the composers Hamza Namira and Waleed Saad from the analysis. This was likely due to the availability of a limited number of data samples for these composers, which could have led to an imbalanced representation.
    \item Maximum of 145 segments (about 5 minutes) for each clip: To ensure consistent segment durations across the dataset, a maximum limit of 145 segments, equivalent to approximately 5 minutes, was set for each clip. This decision aimed to maintain temporal coherence and facilitate accurate analysis and classification.
\end{enumerate}

\par Like the first experiment, the preprocessing steps were nearly the same, as the initial steps in the preprocessing pipeline remained consistent with the first experiment. The MP3 files were converted to the WAV format to ensure compatibility and facilitate further processing. The audio data was then normalized to maintain consistent volume levels across different compositions.

\par Subsequently, the audio compositions were divided into smaller segments, each lasting 5 seconds with a 2-second hop length. This segmentation allowed for a more focused analysis of specific musical parts within the compositions.

\par However, in the second experiment, the padding process for the segments was adjusted. Instead of padding the segments to match a target shape of (20, 938) as in the first experiment, the segments were padded to a target shape of (20, 469). This adjustment ensured that all segments had a consistent shape.

\par By padding the segments, the dimensions of the input data were standardized, enabling consistent processing during model training and evaluation. The padding involved adding zeros to the segments along the second axis (columns) to achieve the desired length.

\par The X\_train, Y\_train, X\_test, Y\_test, X\_val, and Y\_val variables, were saved to be loaded later for model training and evaluation.

\par In the third experiment, we continued using the same dataset as in the previous experiment. However, we decided to leave out the compositions by the seventh composer, Omar Khorshid. This was done to avoid any confusion for the model. Khorshid's tracks are mostly modified versions of existing music pieces, with added instruments like the electric guitar. By not including these tracks, we reduced the chance of the model getting mixed up between Khorshid's unique style and the classical songs of other composers, such as Mohamed Abdelwhab. The preprocessing steps remained consistent with the previous experiments, including the conversion of MP3 files to WAV format, normalization, and segmentation. However, a notable change was made in the feature extraction process. 

\par Instead of utilizing the MFCC (Mel-frequency cepstral coefficients) function, the Mel log spectrogram was employed as a means of capturing the acoustic characteristics inherent in the audio data. The Mel log spectrogram serves as a representation of the audio signal, emphasizing the distribution of frequency components over time.

\par The decision to incorporate the Mel log spectrogram stemmed from the assumption that it could potentially capture features pertaining to the composer's style and musical patterns. By analyzing the representations derived from the spectrogram, the model stood to acquire an enhanced capacity for discerning and distinguishing compositions with greater efficacy.

\begin{figure}[H]
    \centering
    \includegraphics[scale=0.155]{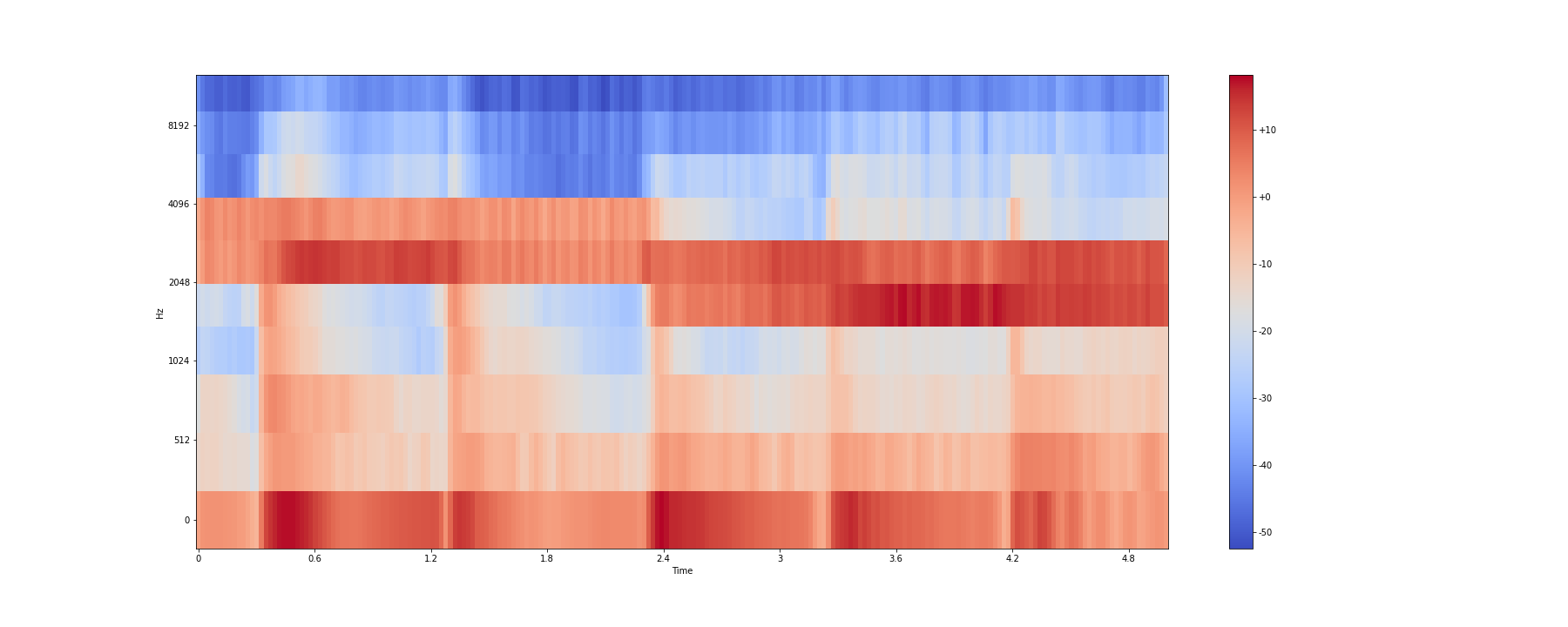}
    \caption{Example of generated Mel-spectrogram}
    \label{fig:VAE3}
\end{figure}

\par Furthermore, it is important to note that the padding step in the preprocessing pipeline adhered to the same target shape employed in the previous experiment, namely (20, 469).

\par In the fourth experiment, in order to enhance the accuracy of the model, it was deemed necessary to augment the dataset by introducing additional data samples. Data augmentation involves creating synthetic data samples by applying transformations or modifications to the existing dataset. These augmentations introduce additional variability and diversity, enabling the model to learn from a more extensive range of examples and improve its ability to generalize.

\par Two augmentation techniques, namely pitch shifting and time stretching, were utilized to enhance the dataset. Pitch shifting was employed by adjusting the pitch of the compositions using pitch shift values of +2 and -2 semitones. This resulted in raising or lowering the overall pitch of the audio samples while maintaining their temporal duration and structure. The model became more robust in recognizing and classifying compositions with different pitch characteristics. Time stretching, on the other hand, involved modifying the temporal duration of the audio samples using time stretch factors of 0.9 and 1.1. This process altered the speed of the compositions, either compressing or expanding their duration while preserving their pitch. 

\par The purpose of applying these augmentation techniques with specific parameter values was to expand the dataset and increase its diversity, enabling the model to learn and generalize better. By training the model on augmented data, it became more adept at accurately classifying compositions based on their respective composers, even when faced with slight variations in pitch and timing.

\par In the fifth experiment, an additional data augmentation technique called frequency masking was introduced to further enhance the model's robustness and variability. Frequency masking involves selectively masking certain frequency bands in the audio data, effectively reducing the presence of specific frequency components. By applying frequency masking, the dataset was augmented with variations in the frequency content of the audio samples. The frequency masking technique works by randomly selecting several frequency bands and masking them by setting their values to zero. This process simulates the presence of missing or distorted frequency components in the audio, which can occur due to various factors such as recording conditions or audio transmission.

\subsubsection{Network Architecture} \hfill
\label{sec: Developing The Model}

\begin{figure}[H]
    \centering
    \includegraphics[scale=0.5]{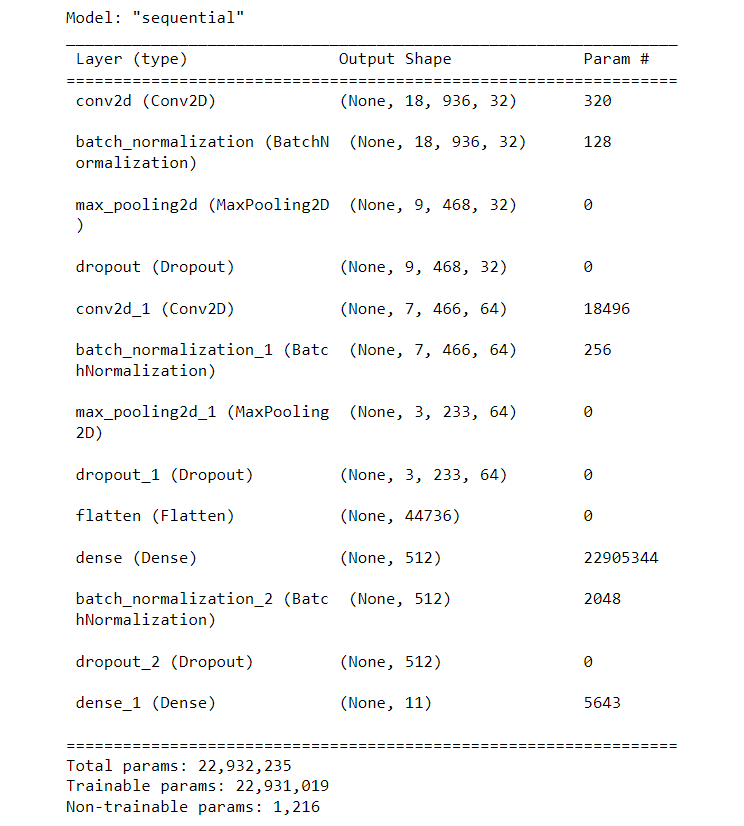}
    \caption{Model Summary for Experiment 1}
    \label{fig:exp1}
\end{figure}

\begin{figure}[H]
    \centering
    \includegraphics[width=0.5\textwidth, height=9.7cm]{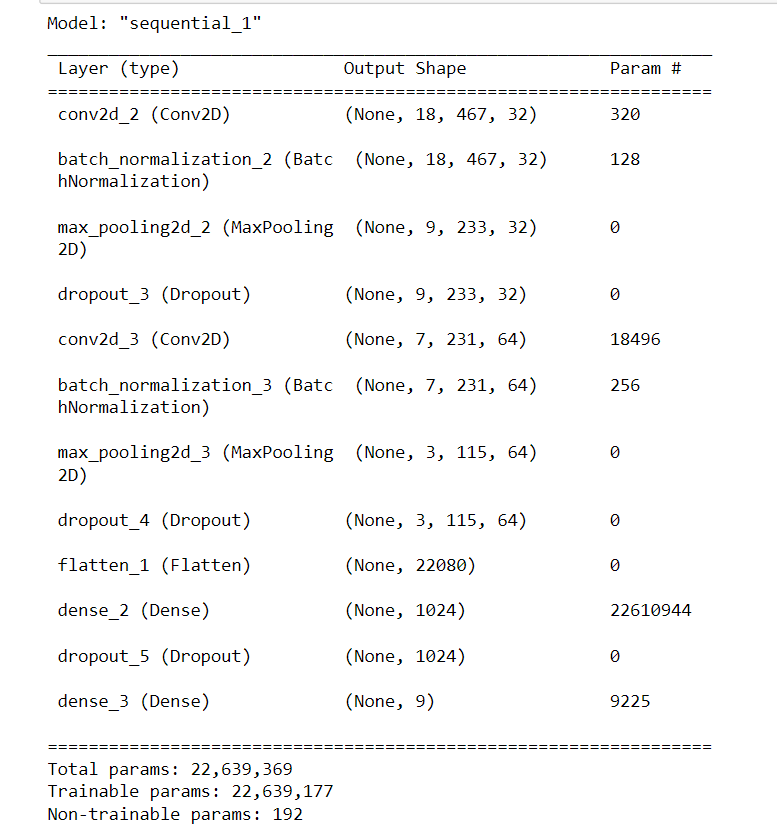}
    \caption{Model Summary for Experiment 2}
    \label{fig:exp2}
\end{figure}

\par In Experiment 1, the model was trained for 30 epochs with a batch size of 32. A sequential model was constructed, as seen in figure \ref{fig:exp1}.
\par In Experiment 2, the model underwent 30 epochs with a batch size of 32. However, the architecture was modified from the first one, as seen in figure \ref{fig:exp2}.
\par
\begin{figure}[H]
    \centering
    \includegraphics[width=0.5\textwidth, height=10cm]{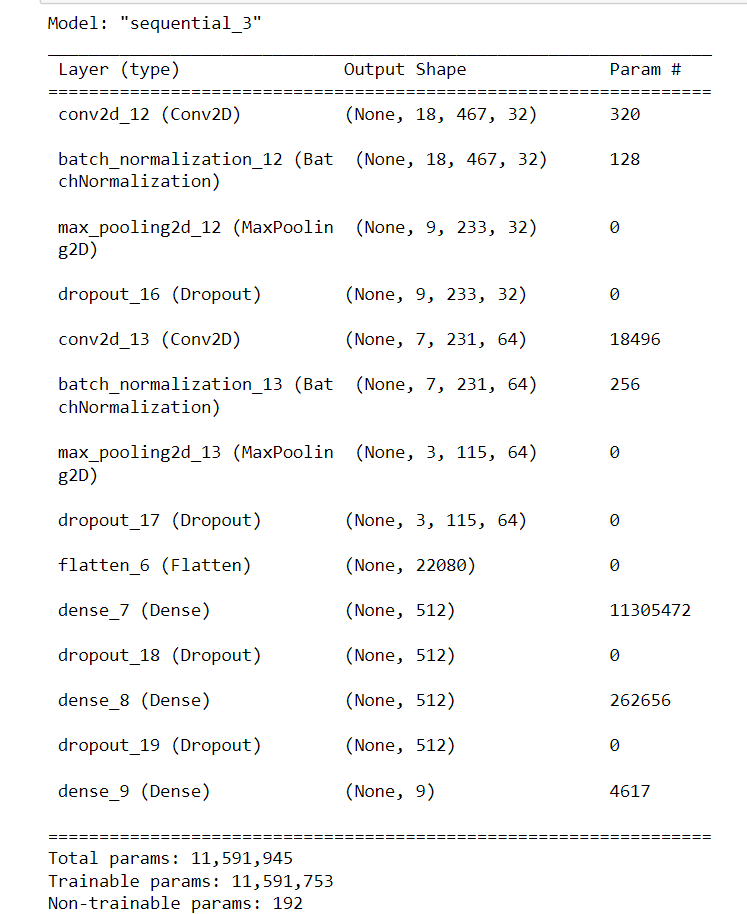}
    \caption{Model Summary for Experiment 3}
    \label{fig:exp3}
\end{figure}

 In Experiment 3, the model architecture was further expanded to increase complexity. The CNN and dense layers were extended as shown in figure \ref{fig:exp3}.

\par Experiments 4 and 5 maintained the same model architecture as Experiment 3 but employed different augmentation techniques in the preprocessing stage. In Experiment 4, the model was trained for 25 epochs with a batch size of 32, while in Experiment 5, the model was trained for 30 epochs with the same batch size.
These experiments allowed us to investigate various model architectures and augmentation techniques to enhance the accuracy and robustness of Arabic music composer classification.

\subsection{Generation Model}
\label{sec: generation}
\subsubsection{Preprocessing}
\label{sec: Preprocessing}

The dataset used for preprocessing consisted of 75 WAV files, which were segmented into a total of 8756 segments. Each audio file was divided into smaller durations of 5.94 seconds, a duration that was determined through multiple iterations and experimentation. The goal was to find a duration that would allow the models to capture sufficient features while maintaining a suitable input size of 256x512.
\\

To determine the optimal duration, the models were tested during each epoch using various durations. The performance of the models was evaluated based on the generated results and the loss errors obtained. Through this iterative process, the duration of 5.94 seconds was found to be effective in capturing relevant audio features while ensuring the models' input requirements were met.

The preprocessing steps involved in preparing these segments for the subsequent model training were as follows:
Segmentation: Each audio file was segmented into smaller durations of 5.94 seconds.

Padding: If necessary, the segmented audio segments were padded with zeros to make them uniform in size. Padding helps maintain consistency in the input dimensions and ensures compatibility with the subsequent processing steps.

Log Spectrogram Extraction: From each segmented audio segment, a log-scaled spectrogram was extracted using the short-time Fourier transform (STFT) method. The STFT divides the audio signal into small overlapping frames and calculates the magnitude of the complex-valued STFT. The magnitude spectrogram was then transformed to the log scale using the logarithmic compression provided by the Librosa library. This conversion enhances the representation of the audio data by emphasizing perceptually relevant features.

Min-Max Normalization: The extracted log spectrograms were normalized using min-max normalization. This process scales the values of the spectrograms to a predefined range, typically between 0 and 1. Min-max normalization ensures that all spectrograms have consistent scaling, enabling the model to learn from a uniform data distribution.

Saving the Preprocessed Data: The preprocessed log spectrograms were saved in the specified paths to facilitate further model training and evaluation. These saved representations serve as the input to the subsequent steps of the music generation pipeline.

By performing these preprocessing steps on the 75 WAV files, the data was effectively transformed and prepared for training the variational autoencoder (VAE) model. The preprocessing pipeline ensures that the audio data is properly formatted, normalized, and ready for subsequent music generation tasks.

\subsubsection{Network Architecture}
\label{sec: Developing The Model}
\par The proposed Arabic music generation model utilizes a Convolutional Variational Autoencoder (CVAE) for the generation process. A CVAE is a type of generative model that combines the power of variational autoencoders with convolutional neural networks to capture the spatial and temporal dependencies in the mel-spectrogram data.

\begin{figure}[H]
    \centering
    \includegraphics[width=0.4\textwidth]{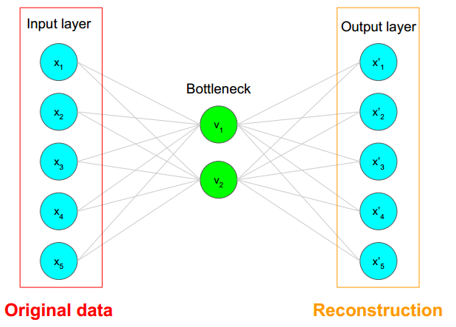}
    \caption{Variational Autoencoder general architecture}
    \label{fig:VAE1}
\end{figure}

\par The encoder component of the CVAE consists of 5 convo
lutional layers, each followed by a ReLU activation function
 and batch normalization. These layers extract hierarchical rep
resentations of the input mel-spectrogram, gradually reducing
 the spatial dimensions. The convolutional filters used in the
 encoder are (512, 256, 128, 64, 32), and the kernel sizes are
 (3, 3, 3, 3, 3). Strided convolutions with strides of (1, 2, 2, 2,
 1) are employed to downsample the feature maps. The output
 of the last convolutional layer is flattened and connected to
 a dense layer with 128 units, which represents the mean and
 variance of the latent space.

\par The decoder component of the CVAE mirrors the architecture of the encoder. It takes the latent space representation as input and gradually upsamples it to generate a mel-spectrogram with the original shape (256, 512). The convolutional layers in the decoder have the same filter sizes, kernel sizes, and strides as the encoder, but in reverse order. 

\par During training, the CVAE aims to minimize two distinct loss functions: the Kullback-Leibler Divergence (KL) loss and the reconstruction loss. The KL loss encourages the learned latent space to follow a prior distribution, typically a standard normal distribution. It measures the divergence between the predicted distribution and the target distribution. The reconstruction loss quantifies the difference between the input mel-spectrogram and the reconstructed output from the CVAE. It ensures that the model can accurately reconstruct the original audio segment.

\begin{equation}
    D_{KL}(N(\mu, \sigma)||(N(0, 1)) = \frac{1}{2} \sum(1+ \log{\sigma^{2}} - \mu^{2} - \sigma^{2})
\end{equation}
where $\sigma$ is the standard deviation of the latent space representation, and $\mu$ is the mean of the latent space representation. The sum is taken over each element of the latent space.

\par To balance the contribution of the two losses, the combined loss is calculated as

\begin{equation}
    Loss = \alpha * RMSE + D_{KL}
\end{equation}
where alpha is a weighting factor. In this Arabic music generation model, the value of alpha is set to 1000000. This value is chosen to prioritize the preservation of the mel-spectrogram structure during reconstruction while also encouraging the latent space to follow the desired distribution.

\par By combining the power of convolutional neural networks, variational autoencoders, and careful design of loss functions, the proposed CVAE-based model for Arabic music generation aims to capture complex patterns and generate novel music sequences in the desired genre.

\subsubsection{Postprocessing}
\label{sec: Postprocessing}

\par After using the CVAE model to generate new mel spectrograms, we need to convert them back to audio. This is done mainly by using the Griffin-Lim algorithm. The Griffin-Lim algorithm is an iterative algorithm that estimates the phase of a spectrogram given its magnitude. The algorithm is described in the paper "Signal estimation from modified short-time Fourier transform"\cite{griffin_lim}. The algorithm is implemented in the Librosa library.
\par The steps for the post-processing are as follows. First, we first reshape the output mel spectrogram to the original shape (removing the additional dimension that we added to fit into the model). Next, we apply denormalization to the mel spectrogram, returning it to its original range. To complete this step, we use the minimum and maximum values of the training data obtained during the preprocessing step.
\par After that, we convert the mel spectrogram to a linear spectrogram using the inverse mel transform. Finally, we apply the Griffin-Lim algorithm to the linear spectrogram using the hop length set earlier to obtain the audio signal. The post-processed signals are saved in a specified directory. These steps are like the inverse of the preprocessing steps that we applied to the data before feeding it into the model. It's necessary to follow this order of steps to get the correct audio signal.
To test the model, we apply the post-processing algorithm to the output of the model and compare it to the original audio signal (post-processing the original input data mel spectrograms).

\section{Results} 
\label{SecResult}
      \subsection{Composers Classification Model}
\par
\begin{figure}[H]
    \centering
    \includegraphics[width=0.44\textwidth, height=6.6cm]{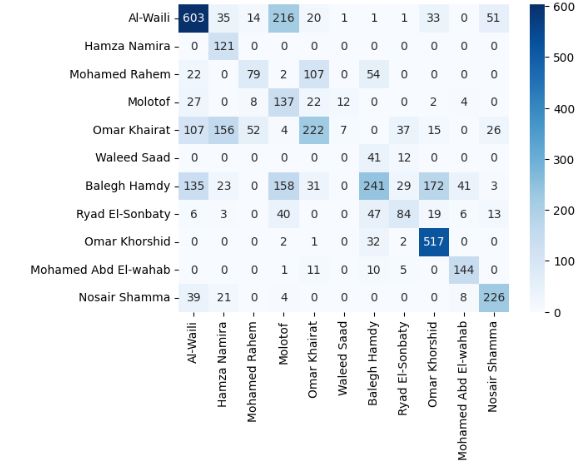}
    \caption{Confusion Matrix for Experiment 1}
    \label{fig:ConfusionMatrix_1}
\end{figure}

 In the first experiment, upon analyzing the confusion matrix, shown in figure \ref{fig:ConfusionMatrix_1}, it became evident that the model struggled to make accurate predictions for most of the composers. This was reflected in the validation accuracy of 0.7370 and the test accuracy of 0.5489.
\par The noticeable difference between the validation accuracy and the test accuracy suggests that the model is overfitting the training data- Overfitting occurs when a model becomes too specialized in learning from the training data and fails to generalize well to unseen data. In this case, the overfitting can be attributed to the uneven distribution of data among the composers, which may have resulted in biased learning.

\par Training accuracy: 0.9936, loss: 5.4657, Test loss: 5.4657, Test accuracy: 0.5489, Validation Accuracy: 0.7370.

\par In the second experiment, the model's complexity was increased by introducing a Conv2D layer with 64 filters and a kernel size of (3, 3) instead of the previous Conv2D layer with 32 filters and the same kernel size.
\par The results showed a decrease in the validation accuracy compared to Experiment 1 as the validation accuracy is 0.5121 while the test accuracy is 0.559. However, it is important to note that the model did not exhibit signs of overfitting. This suggests that the model's performance has improved in terms of generalization, as it is able to handle unseen data more effectively.
\par
\begin{figure}[H]
    \centering
    \includegraphics[width=0.44\textwidth, height=6.6cm]{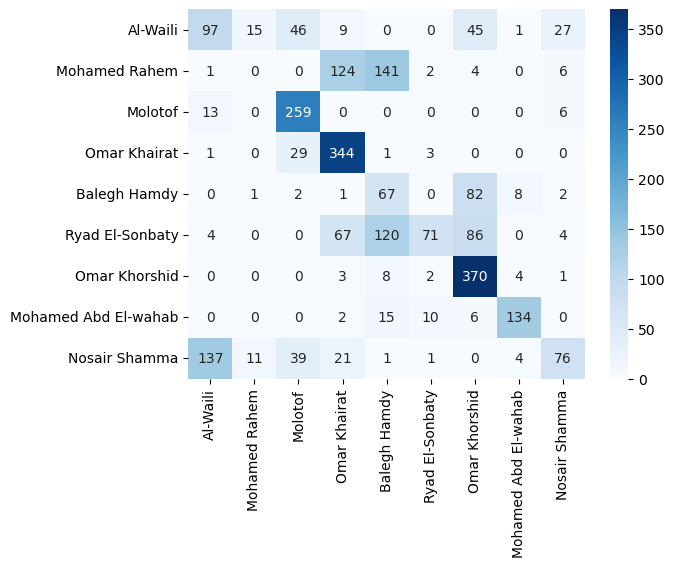}
    \caption{Confusion Matrix for Experiment 2}
    \label{fig:ConfusionMatrix_2}
\end{figure}

 Observing the confusion matrix, figure \ref{fig:ConfusionMatrix_2}, revealed that the model primarily struggled with distinguishing between composers of the same genre as Ryad El-sonbaty with Mohamed Abd El-Wahab and Balegh Hamdy. This observation can be seen as an improvement since it indicates that the model is capable of discerning subtle differences between compositions from similar genres.
\par 
Training  accuracy: 0.9565,
loss: 6.8790,
Test loss: 6.8790,
Test accuracy: 0.5595,
Validation accuracy: 0.6281.
\par In the third experiment, the results showed that the test accuracy improved compared to the previous experiments, achieving a value of 0.7015. However, the validation accuracy was lower at 0.5925, indicating a potential underfitting issue. This suggests that the model's performance could be further improved by increasing the existing data.
\par
\begin{figure}[H]
    \centering
    \includegraphics[width=0.45\textwidth, height=7cm]{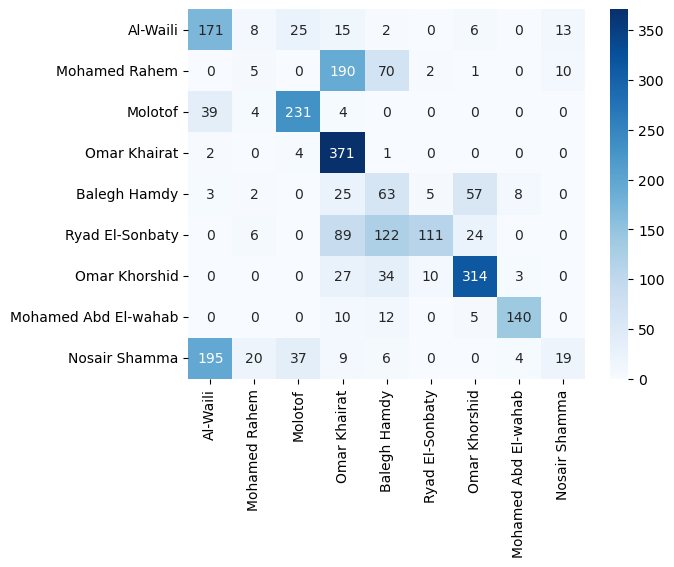}
    \caption{Confusion Matrix for Experiment 3}
    \label{fig:ConfusionMatrix_3}
\end{figure}

 The underfitting observation is supported by the confusion matrix, figure \ref{fig:ConfusionMatrix_3}, which shows that the model struggles to correctly classify compositions from different composers, possibly due to a lack of complexity and capacity to capture subtle differences in musical styles.

\par Experiment 3 demonstrates some improvement in test accuracy, but the presence of underfitting suggests the need for further optimization of the model and increasing the data.
\par Training accuracy: 0.9861,
loss: 1.4560,
Test loss: 1.4560,
Test accuracy: 0.7015,
Validation accuracy: 0.5925.

\par In the fourth experiment, data augmentation techniques were employed to address the underfitting issue encountered in the previous experiment. Two augmentation techniques, namely pitch shifting and time stretching, were applied to modify the audio samples.
\par The results showed a significant improvement compared to the previous experiment. The test accuracy increased to 0.73159, and the validation accuracy improved to 0.6935. This indicates progress in mitigating the underfitting problem and enhancing the model's performance.
\par
\begin{figure}[H]
    \centering
    \includegraphics[width=0.45\textwidth, height=7cm]{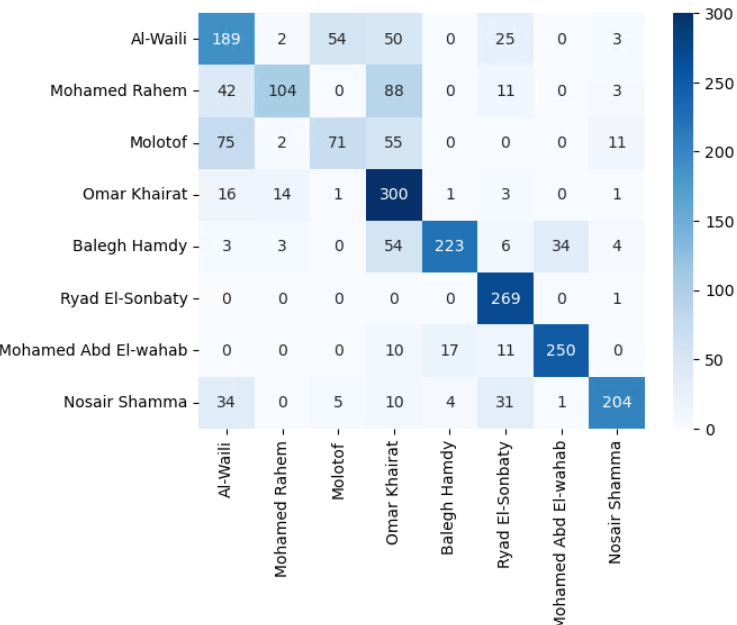}
    \caption{Confusion Matrix for Experiment 4}
    \label{fig:ConfusionMatrix_4}
\end{figure}

 Furthermore, the analysis of the confusion matrix, figure \ref{fig:ConfusionMatrix_4}, showed a notable improvement in the model's ability to distinguish composers with similar genres. For instance, the model significantly reduced its confusion between segments of Ryad El-Sonbaty and Balegh Hamdy, despite the similarity in their musical styles. It accurately predicted most of the segments from Ryad El-Sonbaty, which indicates the model's enhanced capability to differentiate compositions based on their respective composers.

\par 
Training accuracy: 0.9640,
Test loss: 4.6427,
Test accuracy: 0.7316,
Validation accuracy: 0.6935.

\par In the fifth experiment, building upon the promising progress achieved in the previous experiment, an additional augmentation technique which is frequency masking was introduced to further enhance the model's performance. 
\par The inclusion of frequency masking in the augmentation process led to a significant boost in the model's performance. The test accuracy reached a value of 0.814, while the validation accuracy further improved to 0.8458. These results demonstrate the substantial impact of frequency masking on the model's ability to accurately classify compositions based on their composers.
\par
\begin{figure}[H]
    \centering
    \includegraphics[width=0.45\textwidth, height=7cm]{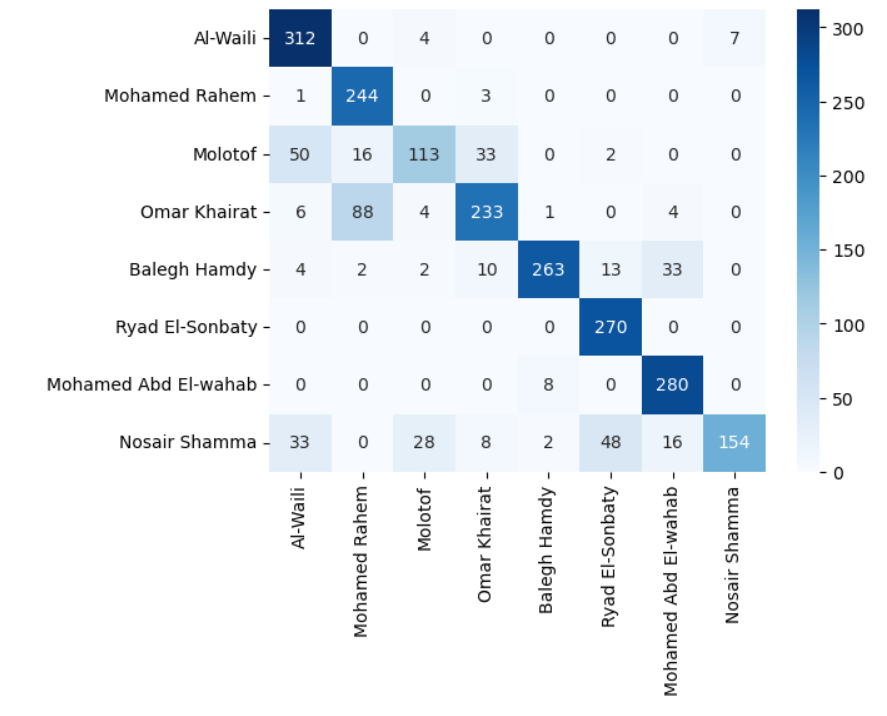}
    \caption{Confusion Matrix for Experiment 5}
    \label{fig:ConfusionMatrix_5}
\end{figure}

 Furthermore, the confusion matrix, figure \ref{fig:ConfusionMatrix_5}, highlighted the diagonal elements, which represent correct predictions, were significantly higher, indicating the model's enhanced capability to classify compositions accurately.
\par 
Training accuracy: 0.9697,
loss: 1.1593,
Test loss: 1.1593,
Test accuracy: 0.8143,
Validation Accuracy: 0.8458.
\subsection{Generation Model}
\par After training our generation model on multiple files, we observed a gradual decrease in the loss function throughout the training process. However, the results were not as promising as we had hoped. Upon listening to the output, it became apparent that while it bore a resemblance to the input, it also contained a slight amount of noise. Despite our rigorous efforts during the preprocessing phase and throughout model training, the outcomes fell short of our expectations.
Our experiments have led us to the conclusion that the variational autoencoder struggled to capture the intricate compositions inherent in our original dataset. To solidify this observation, we recognize the need for additional testing and evaluations. It is noteworthy that this aspect represents our most significant contribution, utilizing the data we originally collected, in the realm of preprocessing and model selection.

\section{Conclusion} 
\label{SecConclusion}

In this study, we investigated the task of classifying different types of composers in the context of Arabic music using deep learning techniques. Through a series of experiments and rigorous data pre-processing, we aimed to achieve the highest possible accuracy in classifying composers based on their musical compositions. Additionally, we explored the generation of Arabic music using deep learning models.

To achieve our classification goals, we conducted five distinct experiments employing various approaches. We identified the best set of pre-processing steps that maximized the accuracy of our models, ensuring optimal representation and encoding of the input data.

For the generation aspect, we experimented with various approaches to generate Arabic music compositions. Our goal was to produce high-quality output that closely resembled the characteristics of the input data. We devoted considerable effort to pre-processing techniques to ensure the generated sound quality. While our current results are a beginning for further improvement, we acknowledge that further refinement and investigation are necessary to achieve even more compelling and accurate generation capabilities.

Overall, this project contributes to the field of Arabic music classification and generation by providing insights into the effective utilization of deep learning techniques. Our experiments highlight the importance of data pre-processing and the impact of different approaches on the accuracy of composers classification. Furthermore, we demonstrate the potential of deep learning models in generating Arabic music compositions. The findings presented here lay the foundation for future research and development in the domain of Arabic music analysis, classification, and generation.

\nocite{hopkins}
\nocite{music}
\nocite{genre}
\nocite{TRAN2021101120}
\bibliographystyle{IEEEtran}
\footnotesize\bibliography{refs}

\begin{thebibliography}{1}
\providecommand{\url}[1]{#1}
\csname url@samestyle\endcsname
\providecommand{\newblock}{\relax}
\providecommand{\bibinfo}[2]{#2}
\providecommand{\BIBentrySTDinterwordspacing}{\spaceskip=0pt\relax}
\providecommand{\BIBentryALTinterwordstretchfactor}{4}
\providecommand{\BIBentryALTinterwordspacing}{\spaceskip=\fontdimen2\font plus
\BIBentryALTinterwordstretchfactor\fontdimen3\font minus \fontdimen4\font\relax}
\providecommand{\BIBforeignlanguage}[2]{{%
\expandafter\ifx\csname l@#1\endcsname\relax
\typeout{** WARNING: IEEEtran.bst: No hyphenation pattern has been}%
\typeout{** loaded for the language `#1'. Using the pattern for}%
\typeout{** the default language instead.}%
\else
\language=\csname l@#1\endcsname
\fi
#2}}
\providecommand{\BIBdecl}{\relax}
\BIBdecl

\bibitem{comp}
D.~Herremans, D.~Martens, and K.~Sörensen, ``Composer classification models for music-theory building,'' Oct. 2015.

\bibitem{kaggleProject}
B.~Holst, ``Composer classification - build model,'' July 2022.

\bibitem{comprehensive}
S.~Ji, J.~Luo, and X.~Yang, ``A comprehensive survey on deep music generation: Multi-level representations, algorithms, evaluations, and future directions,'' 11 2020.

\bibitem{variational}
D.~P. Kingma and M.~Welling, ``An introduction to variational autoencoders,'' 2019.

\bibitem{griffin_lim}
D.~W. Griffin and J.~Lim, ``Signal estimation from modified short-time fourier transform,'' Apr. 1984.

\bibitem{hopkins}
J.~Breebaart and M.~McKinney, \emph{Features for Audio Classification}, 01 2004, vol.~2.

\bibitem{music}
A.~Tikhonov and I.~Yamshchikov, ``Music generation with variational recurrent autoencoder supported by history,'' \emph{SN Applied Sciences}, vol.~2, 12 2020.

\bibitem{genre}
G.~Tzanetakis and P.~Cook, ``Musical genre classification of audio signals,'' \emph{IEEE Transactions on Speech and Audio Processing}, vol.~10, pp. 293--302, 01 2002.

\bibitem{TRAN2021101120}
L.~Comanducci, D.~Gioiosa, M.~Zanoni, F.~Antonacci, and A.~Sarti, ``Variational autoencoders for chord sequence generation conditioned on western harmonic music complexity,'' \emph{EURASIP Journal on Audio, Speech, and Music Processing}, vol. 2023, 05 2023.

\end{thebibliography}
\end{document}